\documentclass[aps,prb,superscriptaddress,showpacs]{revtex4} 

\usepackage{epstopdf}
\usepackage{hyperref}
\usepackage{amsmath}
\usepackage{graphicx}
\usepackage{dcolumn}
\usepackage{bm}
\usepackage{color}

\begin{document}

\title{Thermoelectric efficiency at maximum power in low-dimensional systems}

\author{Natthapon Nakpathomkun}
\affiliation{ 
Department of Physics and Materials Science Institute,
University of Oregon, OR 97403-1274, U.S.A.
}
\author{Hongqi Q. Xu}
\author{Heiner Linke}
\email{heiner.linke@ftf.lth.se}
\affiliation{Division of Solid State Physics and The Nanometer Structure Consortium (nmC@LU), Lund University, P.O. Box 118, 22100 Lund, Sweden
}
\date{\today}

\begin{abstract}
Low-dimensional electronic systems in thermoelectrics have the potential to achieve high thermal-to-electric energy conversion efficiency. A key measure of performance is the efficiency when the device is operated under maximum power conditions.  Here we study the efficiency at maximum power of three low-dimensional, thermoelectric systems: a zero-dimensional quantum dot (QD) with a Lorentzian transmission resonance of finite width, a one-dimensional (1D) ballistic conductor, and a thermionic (TI) power generator formed by a two-dimensional energy barrier. In all three systems, the efficiency at maximum power is independent of temperature, and in each case a careful tuning of relevant energies is required to achieve maximal performance. We find that quantum dots perform relatively poorly under maximum power conditions, with relatively low efficiency and small power throughput. Ideal one-dimensional conductors offer the highest efficiency at maximum power (36\% of the Carnot efficiency). Whether 1D or TI systems achieve the larger maximum power output depends on temperature and area filling factor. These results are also discussed in the context of the traditional figure of merit $ZT$.
\end{abstract}


\maketitle
\section{Introduction}
There is broad, current interest in developing high-performance thermoelectric (TE) materials for the generation of electric power from heat sources such as waste heat. Of particular interest are nanostructured materials for two primary reasons: first, a high density of interfaces, on a length scale comparable to the phonon mean free path, can be used to reduce parasitic heat flow carried  by the crystal lattice \cite{Harman02,Bottner06,Snyder08}. Second, a reduced dimensionality of the electronic system, achieved by band engineering or by nanostructuring on the scale of the electron wave length, can be used, in principle, to optimize a material's electronic properties. Specifically, sharp features of the electronic density of states (DOS) can increase the thermopower (Seebeck coefficient) $S$, which is a measure of the average kinetic energy of mobile electrons relative to the chemical potential, and can thus increase the power factor \cite{Hicks93a,Hicks93b,Mahan96,Heremans08, KimR09} $S^2\sigma$, where $\sigma$ is the electric conductivity.

To characterize a TE material, one commonly uses the figure of merit $ZT=S^2\sigma T/(\kappa_e+\kappa_{l})$ which is a function of the power factor and of the electron's ($\kappa_e$) and lattice's ($\kappa_{l}$) contributions to the parasitic thermal conductivity. $ZT$ is closely related to the achievable efficiency of thermal-to-electric energy conversion. A delta-function shaped DOS maximizes $ZT$ \cite{Mahan96}, and can be used to establish reversible thermal-to-electric energy conversion at the thermodynamically maximal Carnot efficiency $\eta_C = \Delta T/T_H$ \cite{Humphrey02,Humphrey05,ODwyer06, Esposito09}, where $\Delta T = T_H - T_C$ is the temperature difference between a hot (H) and cold (C) electron reservoir at temperatures $T_H$ and $T_C$, respectively (Fig.~\ref{diagsystem}). The development of strategies to realize operation near $\eta_C$  \cite{Humphrey05b,ODwyer06} helps understanding the sources of irreversible conversion losses and is thus of both fundamental and practical interest. However, efficiency near $\eta_C$ requires near-reversible operation, a limit where the power output necessarily goes to zero. For practical applications it is thus of greater interest to understand the relationship between efficiency and power production. The relevant fundamental efficiency limit in this context is the Curzon-Ahlborn limit 
\begin{equation}
	\eta_{CA}=1-\sqrt{T_C/T_H} \approx \frac{\eta_C}{2}+\frac{\eta_C^2}{8}+\mathcal O(\eta_C^3)+\dotsc,
\end{equation}
which is the thermodynamically maximum efficiency of a heat engine operating under conditions where the maximum power output is maximized \cite{Curzon75}. Esposito \emph{et~al.} \cite{Esposito09, Esposito10} recently showed that zero-dimensional quantum dots (QDs) with a delta-like transmission function, tuned to produce maximum power, have a TE conversion efficiency that matches the Curzon-Ahlborn (CA) limit up to quadratic terms in $\eta_C$. However, limiting the analysis to a delta-like transmission function sets the QD's power output to negligible values. Finite broadening of the transmission function is needed to reach full maximum power, but is associated with a reduction in efficiency \cite{Humphrey02}. Prior work has explored electron energy filters with a digital (on-off) energy dependence as a function of filter width \cite{Humphrey02,  Humphrey03}. The properties of a realistic quantum-dot energy filter based on resonant tunneling have not yet been explored.

Here we ask: which low-dimensional electronic system yields the highest thermoelectric (TE) efficiency under maximum power conditions? Specifically, we address three idealized systems: first, a zero-dimensional quantum dot (QD) with finite, Lorentzian broadening of the transmission resonances, allowing for maximized power production; second, an idealized quantum wire (1D), and third, a two-dimensional, thermionic (TI) energy barrier embedded into a bulk material with a lower bandgap. Our goal is to establish which of these electronic systems fundamentally offers the best trade-off between thermoelectric power and efficiency. For this purpose, we assume ideal electronic properties such that our conclusions become independent of specific material parameters, such as mean free paths, with the only exception of the effective mass in TI system.  For the TI system, we also focus on the properties of the barrier itself, and do not consider the TE properties of the 3D host material. In addition, we do not include the phonon contribution to heat flow in our calculations: it is external to the electronic system and can be added to the analysis by considering the associated parasitic heat flow.

Our work is complementary to recent work by Kim \emph{et~al.} \cite{KimR09} who compared Seebeck coefficient $S$ and power factor in generic 1D, 2D, and 3D electronic systems. That study found that, when tuning each system to its maximum power factor, $S_{1D}>S_{2D}>S_{3D}$, and that 1D systems also have the highest power factor per mode. Here, we add the QD and TI systems to the comparison, and do not focus on the power factor, which is evaluated under open-circuit conditions. Efficiency at maximum power, in contrast, is achieved under finite current conditions at a specific operating point, and we evaluate the actual power and efficiency at that operating point. 

In the following, in Section II, we will first describe our technical approach, and calculate for each system the operating conditions (in terms of bias voltage and position of the chemical potential) that produce, at given $\Delta T$, maximum power, maximum efficiency ($\eta_{max}$)	as well as maximum efficiency at maximum power ($\eta_{maxP}$), and we discuss the mechanism that reduces $\eta_{max}$ below the CA value for each system. For all three systems $\eta_{maxP}$ is found to be independent of temperature. We then compare our results across the three systems and place our results into the context of the traditional figure of merit, $ZT$.

\section{Modeling details}

\begin{figure}[htbp]
	\centering
	\includegraphics[scale=0.75]{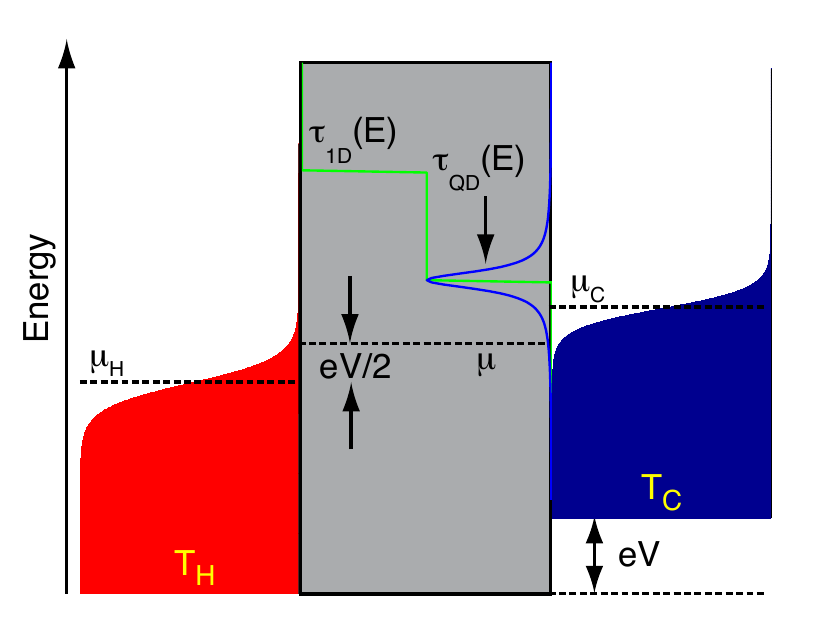}
	\caption{ \label{diagsystem}(Color online) The basic setup considered here consists of a device described by its transmission function $\tau(E)$ ($\tau_{QD}$ and $\tau_{1D}$ are sketched as examples), with contact leads that act as the hot and cold electron reservoirs. A bias voltage, $V$, is applied symmetrically with respect to the average chemical potential $\mu$, which can be tuned relative to the transmission function, using a gate voltage.}
\end{figure}

Figure \ref{diagsystem} shows the generic device configuration considered here.  A cold (C) and hot (H) electron reservoir each obey a Fermi Dirac contribution:
\begin{equation}
	f_{H/C}=\left[1+\exp\left(\frac{E-\mu_{H/C}}{k_BT_{H/C}}\right)\right]^{-1}.
\end{equation}
For simplicity in the calculations, we assume that the voltage difference $V$ is applied symmetrically across the device ($\pm V/2$) such that the electrochemical potentials in the hot and cold electron reservoirs are given by $\mu_{H/C} = \mu \pm eV/2$, where $\mu$ is the chemical potential of the system at equilibrium ($V=0$).

The two reservoirs are connected by a device that acts as an energy-selective filter described by its transmission function $\tau(E)$, which depends on the dimensionality of the device. In an experiment, a gate can be used to tune $\mu$ relative to the device's transmission function, which we here assume to be independent of the gate voltage.

To generate thermoelectric power, the transmission function must be positioned such that high-energy electrons from the hot side can move, against applied bias voltage, into empty states on the cold side, but (ideally) not vice versa. Power is increased by large net electron flow from hot to cold, regardless of electron energy. Efficiency is optimized when this is achieved at minimal heat transport, for example by suppressing the flow of electrons with relatively large kinetic energy (energy filtering). 

In the following, we explore this balance of power and efficiency optimization, as a function of $\mu$ and $V$, for our three different low-dimensional systems.  All numerical calculations are performed for a fixed ratio of $\Delta T/T_C = 0.1$. Note that, in general, nonlinear effects are important, and results will depend on $\Delta T$ for large $\Delta  T/T$ \cite{Humphrey06}.

\subsection{Quantum Dot}

A quantum dot is a zero-dimensional system with well-defined energy levels defined by a combination of size quantization and Coulomb-repulsion effects. One implementation is a double-barrier resonant tunneling quantum dot embedded in a nanowire \cite{Bjork04, ODwyer06}, with the leads corresponding to the electron reservoirs in Fig.~\ref{diagsystem}. 
 
Here we consider the use of a dot's discrete energy spectrum as an energy filter: the dot transmits electrons with energy corresponding to an energy level $E_0$ inside the dot via resonant tunneling. We assume that transport through the dot is elastic, and that the energy level separation $\Delta E $ is much larger than the thermal energy $kT$, such that only the resonance located at $E_0$ contributes to electron transport. 

Thermoelectric transport through a quantum dot with one dimensional leads can be described by the Landauer formalism,
\begin{equation} 
	\label{eq:current}
	I=\frac{2e}{h}\int (f_H-f_C)\tau_{QD}(E)dE.
\end{equation}
The transmission function $\tau_{QD}$ of a single energy level at energy $E_{0}$ is approximated by a Lorentzian function \cite{Datta95} as 
\begin{equation}
	\tau_{QD}(E)=\frac{(\Gamma/2)^2}{(E-E_0)^2+(\Gamma/2)^2}
\end{equation}
where $\Gamma$ is the full width at half maximum of the transmission function.

\begin{figure}[htbp]
	\centering
	\includegraphics[scale=0.75]{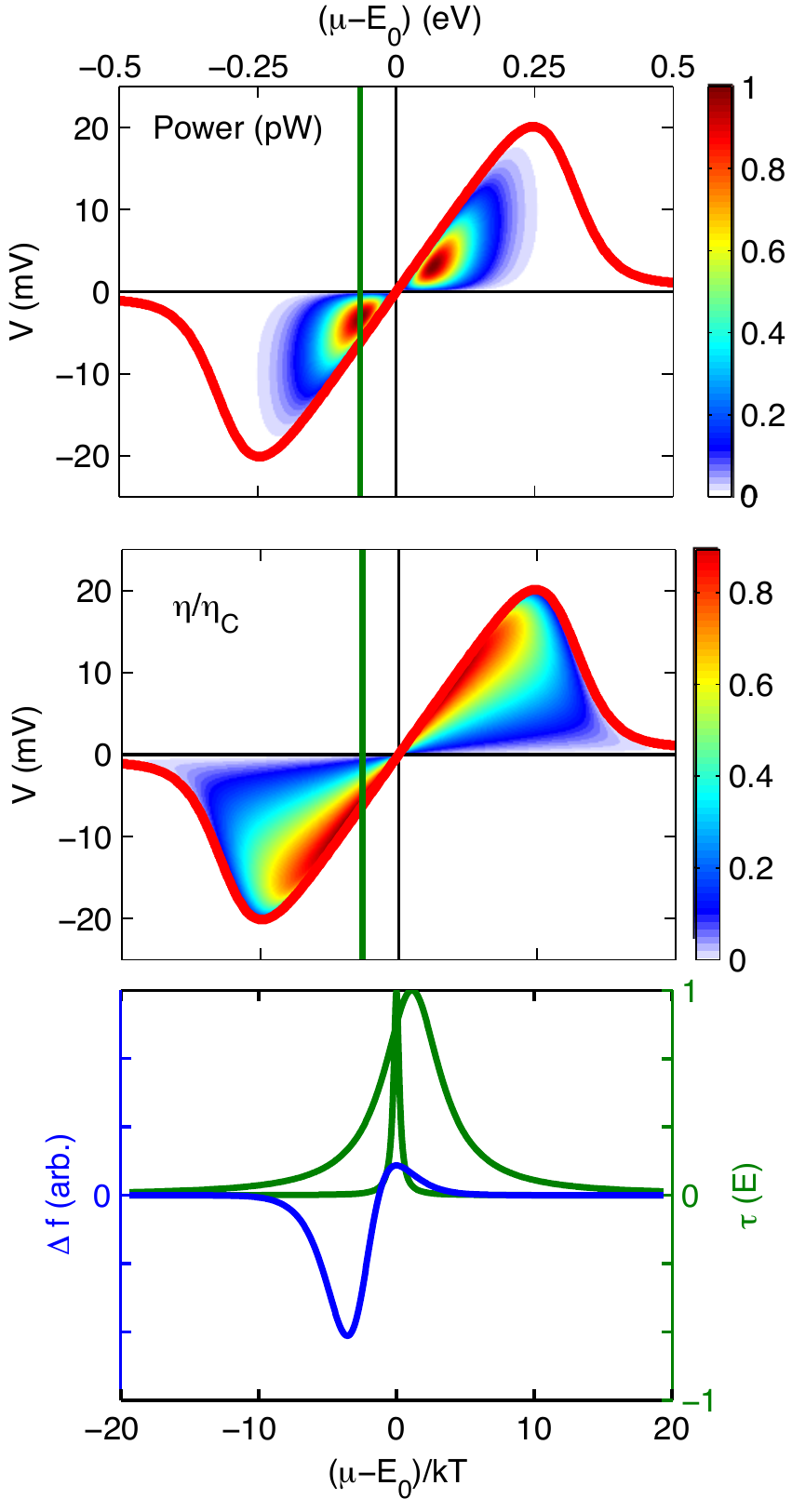}
	\caption{\label{0D_300K}(Color online) (a) Power  and (b) efficiency normalized by Carnot efficiency, of a QD as a function of bias voltage $V$ and average chemical potential $\mu$, for $T_C = 300$ K, $T_H = 330$ K ($\Delta T/T_C = 0.1$) and $\Gamma=0.01kT$ . The open circuit voltage, $V_{oc}$ is highlighted in red (peak $V_{oc}$ corresponds to $S \approx 2$ meV/K). The system works as a generator when the bias is between zero and $V_{oc}$. The vertical green line indicates the  $\mu$ where maximum power occurs. (c) Current through a QD is the integral over the product of $\tau_{QD}$ (green) (Eq.~\ref{eq:current})  and $\Delta f = (f_H-f_C)$, shown here in blue, using the $\mu$ and $V$ that result in $P_{max}$. Two transmission widths, $\Gamma = 0.5 kT$ and $5kT$ are plotted here in the approximate position where maximum  power would be achieved.} 
\end{figure}

The net heat flux ($\dot{Q}_H$) out of the hot reservoir is given by,
\begin{equation}
	\label{eq:heatflux}
	\dot{Q}_H = \frac{2}{h}\int (E-\mu_H) (f_H-f_C) \tau(E) dE,
\end{equation}
and electric power output and thermoelectric efficiency are given by $P=IV$ and $\eta=P/\dot{Q}_H$, respectively. In Fig.~\ref{0D_300K}(a) and \ref{0D_300K}(b), we show calculations of each of these quantities as a function of $V$ and $\mu$. Data are shown only where electric power is produced by the dot, and efficiency is normalized by $\eta_C$. The red line, where the current driven by the temperature gradient is canceled by that driven by the voltage difference, represents the open circuit voltage ($V_{oc}$) as a function of ($\mu - E_0$). The line shape of $V_{oc}$ is determined by the shape and width of the transmission function  \cite{Persson10} , which results from the coupling strength to the leads, contributing tunneling processes \cite{Scheibner07}, as well as by $\Delta E$ when $\Delta E \approx kT$  \cite{Persson10} (this case is not considered here). Inside the pocket defined by $V_{oc}$, the system acts as a power generator. For example, when $\mu$ is located within a few $kT$ below a resonance of $\tau_{QD}$ (Fig.~\ref{diagsystem}) at $E_0$, hot electrons move from occupied states on the hot side to empty states on the cold side, even in the presence of a small counter bias.

Figures \ref{0D_300K}(a) and \ref{0D_300K}(b) can be used to read out the operating points for maximum power and maximum efficiency, which occur at different sets of $\mu$ and $V$: efficiency is highest near $V \approx V_{oc}$, whereas power is optimized closer to $\mu \approx E_0$, and at intermediate $V$. This can be easiest understood by considering a narrow transmission function $(\Gamma << kT)$: in this case, the efficiency is maximal when the resonance is positioned at the energy where $(f_H-f_C) = 0$ (see Fig. \ref{0D_300K}(c)), because here transport is reversible \cite{Humphrey02}, and this equilibrium condition also defines $V_{oc}$. Power is maximized (i) when $\mu_H$ is lined up within a $kT$ or so below $E_0$, such that current is maximized, and (ii) when $V \approx V_{oc}/2 $, in order to optimize the product $P = IV$.

Note that the quantitative results presented here are for $T_C = 300$ K and  \mbox{$T_H = 330$ K}. Whereas these are very high temperatures in the context of quantum-dot transport experiments, which are typically performed around 1K, quantitative data at this temperature are easier to interpret in the context of thermoelectric applications.  The results are valid as long as $kT \ll \mu, \Delta E$ and in the absence of electron-phonon interactions, and for a QD we find that the thermoelectric properties scale as $\Gamma/kT$.
 
\begin{figure}[htbp] 
   \centering
   \includegraphics[scale=0.75]{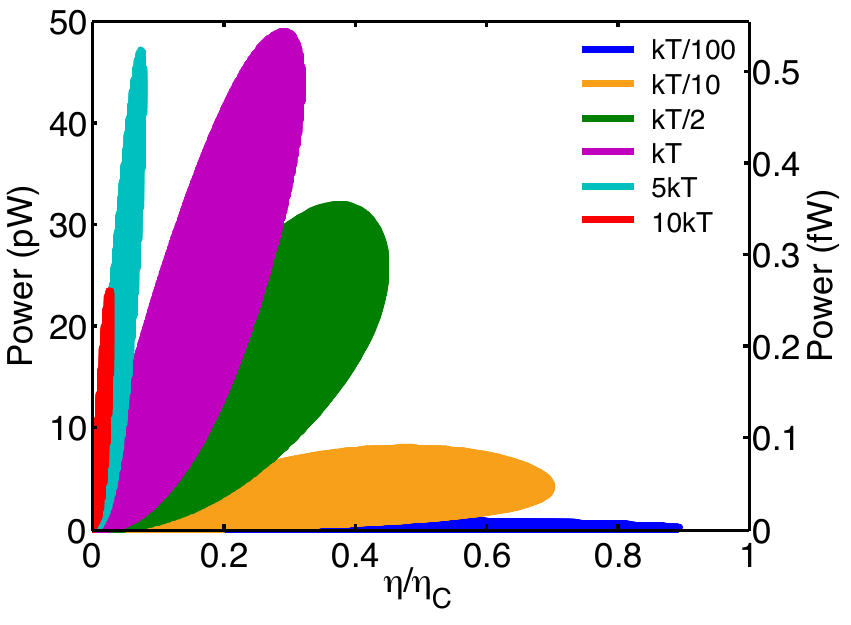} 
   \caption{\label{qd various width}(Color online) Power versus normalized efficiency of a quantum dot for various $\Gamma$ as indicated. The left vertical axis shows data for $T_C = 300 $ K and $T_H = 330$ K whereas the right vertical axis shows the power for $T_C = 1$ K and $T_H = 1.1$ K. The shape of the curves doesn't depend on temperature.}
\end{figure}

We now discuss the influence of the resonant level width $\Gamma$ on the QD power and efficiency. We begin by tracing both power and efficiency along constant $\mu$, that is, along a vertical line in Figs.~\ref{0D_300K}(a) and \ref{0D_300K}(b). For each operating point ($\mu$, $V$), we then graph the power at that point as function of the corresponding efficiency.  The result is a loop, and all loops for all $\mu$ fill up a region in $(\eta, P)$ space, whose shape and position depend on $\Gamma$, as shown in Fig.~\ref{qd various width}. 

\begin{figure}[htbp]
	\centering
	\includegraphics[scale=0.75]{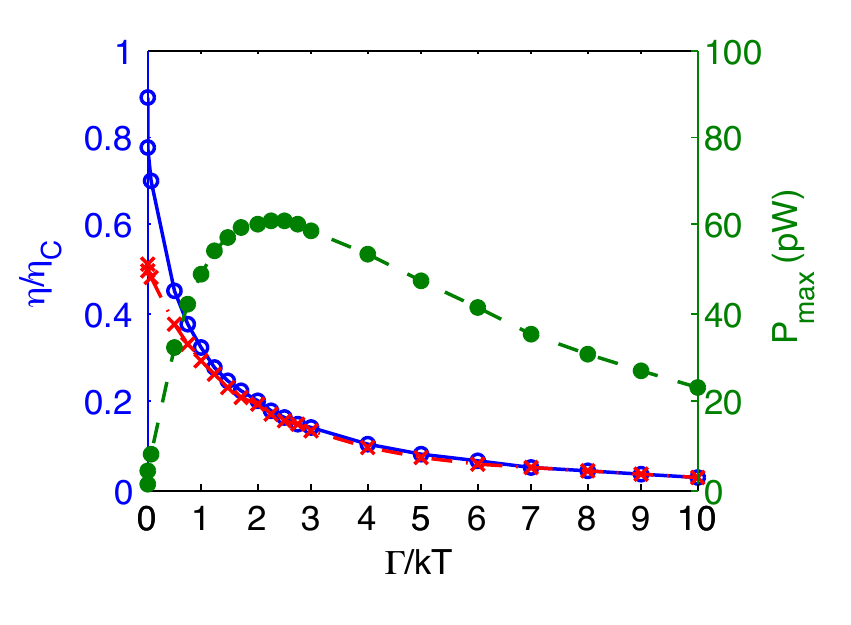}
	\caption{\label{qd eff pmax vs fwhm}(Color online) $\eta_{maxP}$ (red, crosses) and $\eta_{max}$ (blue, open dots), both normalized by Carnot efficiency, and maximum power (green, full dots) of a quantum dot as a function of $\Gamma/kT$ for $T_C = 300$ K and $T_H = 330$ K. Maximum power peaks around $\Gamma/kT = 2.25$. Efficiency at maximum power $\eta_{maxP}$ approaches $\eta_{CA} =$ 51\% for small $\Gamma$ and is always smaller than $\eta_{max}$.}
\end{figure}

For a very small $\Gamma$ (see $\Gamma = 0.01kT$ in Fig.~\ref{qd various width}), $\tau_{QD}(E)$ approaches a delta function, and the maximum efficiency approaches the Carnot limit, as expected \cite{Humphrey02}, but the maximum power is very small. A larger $\Gamma$ allows more electrons to contribute to power generation, and power increases until it reaches a peak value at $\Gamma \approx 2.25kT$, whereas the maximum efficiency monotonically decreases with increasing $\Gamma$ (Fig.~\ref{qd eff pmax vs fwhm}). To understand this behavior, consider Fig.~\ref{0D_300K}(c), which shows examples for $\tau_{QD}(E)$ overlayed onto $\Delta f = (f_H-f_C)$ at finite bias voltage. The integral over the product of both functions determines the current (Eq.~\ref{eq:current}). Electron transport is reversible (a requirement for Carnot efficiency) only where $\Delta f = 0$ \cite{Humphrey02}. Away from this energy, electrons either travel in the wrong direction (where $\mu < E_n$ in Fig.~\ref{0D_300K}(c)), or they carry excessive kinetic energy, ÒwastingÓ heat in electricity production. This explains the drop in efficiency for increasing $\Gamma$. As $\Gamma$ increases, the Lorentzian-shaped $\tau_{QD}(E)$ samples over a wide energy range, and at some point it is not possible to increase $\Gamma$ without also substantially sampling regions where  $\Delta f < 0$ , such that power decreases beyond an optimal value for $\Gamma$. 

\subsection{Nanowire}

Here we consider an ideal 1D-electronic system with a width comparable to the Fermi wavelength ($\lambda_F$), and with a length shorter than the electron mean free path, such as a quantum point contact  defined in a two-dimensional electron gas \cite{vanWees88, Wharam88}. As a result of confinement, the electron energy is quantized in the two lateral dimensions, but can assume any value along the transport direction $x$: 
\begin{equation}
E(x,y,z)=E_n(y,z)+\frac{\hbar^2k_x^2}{2m^*},
\end{equation}
where $n \geq  1$ is the integer subband number and $m^*$ is the electron effective mass.

Using a step-function shaped transmission function
\begin{equation} \label{eq:tau 1D}
	\tau_{1D}(E)=\sum_{n=1}^{\infty} \Theta(E-E_n)
\end{equation}
in the Landauer formalism, we can calculate the current and heat flux out of the hot side using Eqs.(\ref{eq:current}) and (\ref{eq:heatflux}), respectively \cite{Streda89}. For a more realistic description of a quantum-point contact, one can use the analytic transmission function of a saddle-point potential, \mbox{$\tau(E)=\sum_{n=1}^\infty \left [1+\exp(\frac{2\pi(E_n-E)}{\hbar\omega_x})\right ]^{-1}$ with $E_n=V_0+(n-\frac{1}{2})\hbar\omega_y$}, where $V_0$ is the height of the saddle, and   the longitudinal and  lateral curvatures of the saddle-point potential are characterized by the angular frequencies, $\omega_x$ and $\omega_y$, respectively. Note that as $\omega_x/\omega_y \rightarrow 0$, the saddle-point transmission approaches the step-function transmission. Maximum power and efficiency from the two transmission functions, for $\hbar\omega_y >> kT$,  agree within 2\% for $\hbar\omega_x < kT $ (Table \ref{step vs saddle}), and in the following we use Eq.~(\ref{eq:tau 1D}). 

\begin{table}[htdp]
	\begin{center}
	\begin{tabular}{|c|c|c|c|} 
\hline
$\tau$     				& $p_{max}$ (fW)  & normalized EMP	& $\omega_x/\omega_y$\\
\hline
step fn. 				& 1.8180 	& 36.21 \%	&  \\
$\hbar\omega_x=0.01kT	$	& 1.8176	& 36.21\%		& $1.7e-4$ \\
$\hbar\omega_x=0.1kT$		& 1.8172 	& 36.21\%		& $1.7e-3$ \\
$\hbar\omega_x=kT$ 		& 1.7746	& 35.66\%		& $1.7e-2$ \\
\hline
	\end{tabular}
\end{center}
\caption{Comparing the maximum power and normalized efficiency at maximum power for step function and saddle-point transmission function. $\hbar\omega_y = 5$ meV, $T_C = 1$ K, and $T_H = 1.1$ K.}
\label{step vs saddle}
\end{table}

\begin{figure}[htbp]
	\centering
	\includegraphics[scale=0.75]{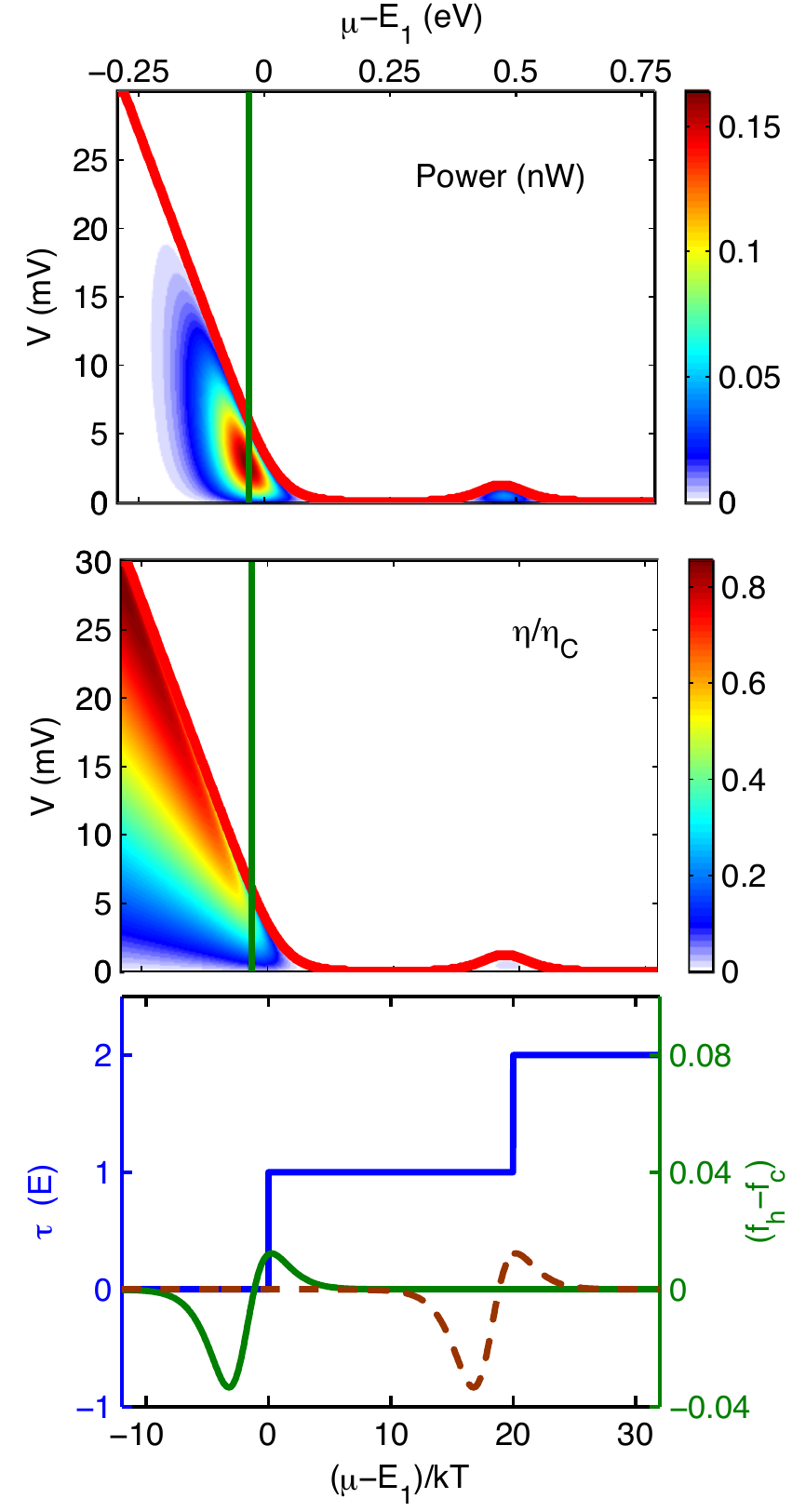}
	\caption{\label{1D_300K}(Color online) Thermoelectric performance of a 1D conductor. (a) Power (nW) and (b) normalized efficiency, $\eta/\eta_C$, near the first ($E_1$) and second($E_2$) subband edges for $T_H = 330$ K and $T_C = 300$ K. The full green line indicates the $\mu$ for maximum power output. (c) Transmission function $\tau_{1D}$ (Eq.~(\ref{eq:tau 1D}), blue line) and $\Delta f = f_H-f_C$ near the onset of the lowest (green) and second subbands (brown) with $V=-2$mV and $(E_2- E_1) = 0.5$ eV.}
\end{figure}

The thermopower of a quantum-point contact is known to strongly depend on the number of occupied subbands \cite{Streda89,Molenkamp90,vanHouten92}, and as one may expect, we find the same for power and normalized efficiency as shown in Figs.~\ref{1D_300K}(a) and \ref{1D_300K}(b). The reason for why a wire with one occupied subband ($n=1$) performs comparatively much better than the one with two occupied subbands ($n=2$) is indicated in Fig.~\ref{1D_300K}(c):  because transmission below the first subband is zero, it is possible to tune $\mu$ to a value just below $E_1$, such that $\Delta f $ (solid green line) allows electron flow only from hot to cold. This is not possible near the onset of the second subband ($E_2$) where $\Delta f $ (solid brown line) will always allow parasitic electron back-flow carried by the first subband, reducing current, power and efficiency. In the following, we therefore focus on wires with $\mu$ in the vicinity of $E_1$ and $(E_2-E_1) \gg kT$ .

Similar to the QD case, the power of a 1D power generator is largest within a few $kT$ of $E_1$(indicated in Fig.~\ref{1D_300K}(a)), allowing significant thermal excitation from $\mu_H$ into the first 1D subband, and for $V \approx 0.5 V_{oc}$, where the product $IV$ is maximized. 

\subsection{Thermionic Power Generator}

A thermionic refrigerator or power generator is based on an energy barrier that preferentially transmits electrons with high kinetic energy. Whereas the original concept \cite{Mahan94} considered use of a vacuum diode, Shakouri and Bowers \cite{Shakouri97} introduced the use of semiconductor heterostructures with energy band-offsets more suitable for room temperature operation with high power density. It is important to note that a 2D energy barrier embedded into a 3D bulk material filters electrons with respect to their cross-plane momentum, and not actually with respect to their total energy \cite{Humphrey05, Vashaee04}. Here we assume conservation of in-plane momentum and a sufficiently thick barrier such that tunneling is suppressed \cite{Mahan98}, and we assume ballistic and elastic transport across the barrier. 

\begin{figure}[htbp]
	\centering
	\includegraphics[scale=0.75]{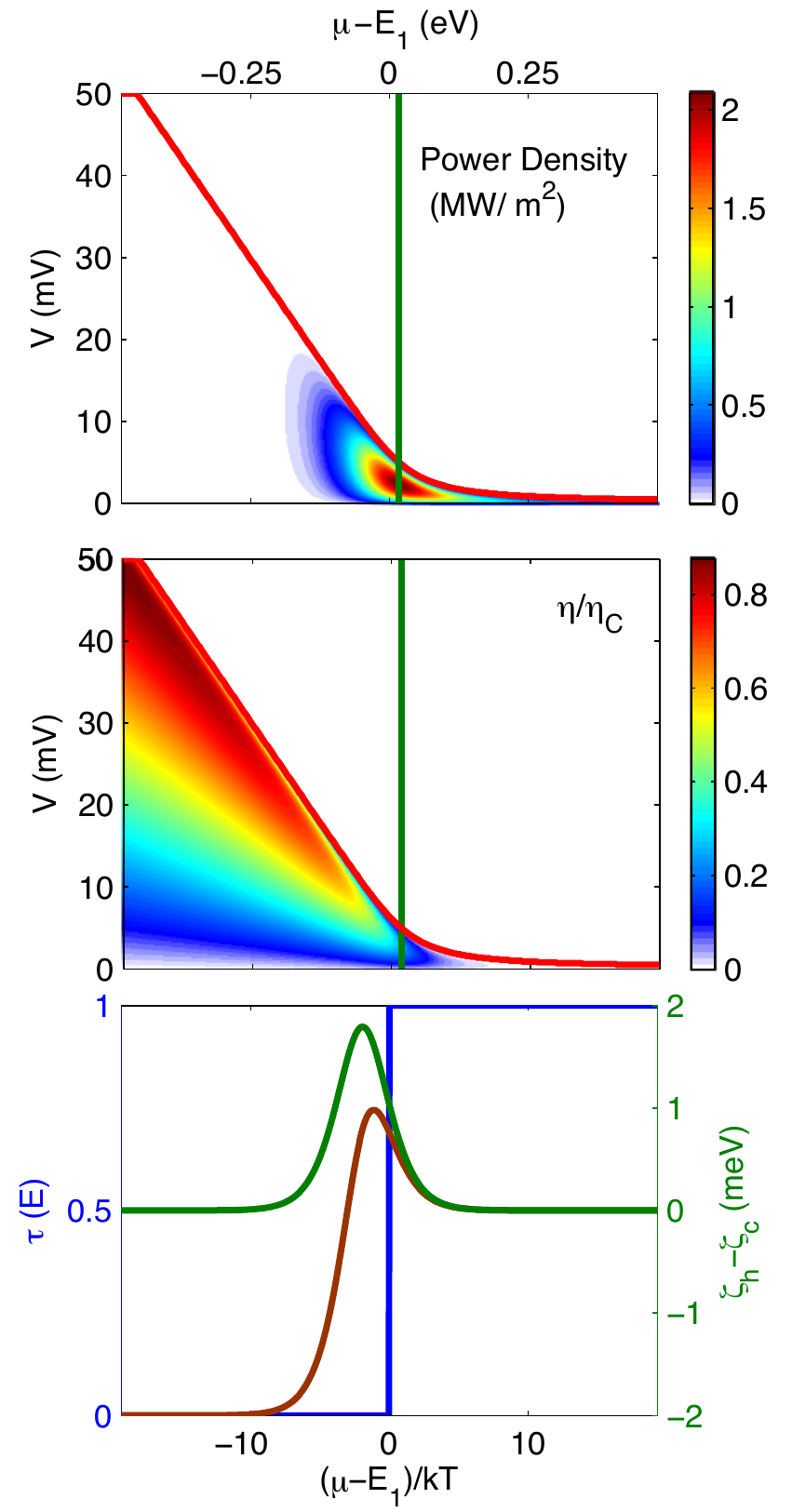}
	\caption{\label{2D_300K}(Color online) (a) Power density (in MW/m$^2$) and (b) normalized efficiency $\eta/\eta_C$ for a thermionic (TI) energy barrier as a function of $V$ and $\mu$ for $T_H = 330$ K and $T_C = 300$ K. (c) Transmission function $\tau_{TI}(E)$ (Eq.~\ref{eq:tau TI}) and $\Delta \zeta =\zeta_H-\zeta_C > 0$ (Eq.~\ref{eq:zeta}) for $V$ = 0 (green) and for  $V$ = -2 mV  (brown) and for $\mu-E_1$  = 0.05 eV  where maximum power is produced. The choice of $\mu > E_1$ ensures that all of the region where $\Delta \zeta =\zeta_H-\zeta_C > 0$ contributes to current flow.}
\end{figure}

Using the Tsu-Esaki formula \cite{Ferry97}, the current density is given by
\begin{equation}\label{TsuEsaki}
J= \frac{m^*e}{2 \pi^2 \hbar^3}\int [\zeta_H-\zeta_C ] \tau(E_x) dE_x,
\end{equation}
where 
\begin{equation}\label{eq:zeta}
\zeta_{H/C} = kT_{H/C} \log\left[1+\exp \left(-\frac{E_x-\mu_{H/C}}{kT_{H/C}}\right)\right],
\end{equation}
and $m^*$ is the effective mass of electrons.  Disregarding tunneling, the transmission function (Fig.~\ref{2D_300K}(c)) is given by 
\begin{equation}
	\tau_{TI}(E_x) = \Theta(E_x-E_b)
	\label{eq:tau TI}
\end{equation}
with $E_b= (\hbar k^\prime_x)^2/2m^*$ and $\hbar k^\prime_x$ is the minimal cross-plane momentum needed to cross the barrier.

The heat flux out of the hot side per unit area is given by
\begin{equation}
 \dot{q}_H=\frac{m^*}{2\pi^2\hbar^3} \int [\phi_H\zeta_H-\phi_C\zeta_C]\tau(E_x)dE_x,
\end{equation}
where $\phi_{H/C}=(E_x+kT_{H/C}-\mu_H)$
and  power density (per unit area) and efficiency are obtained from ${\mathcal P_{TI}}=JV$ and $\eta={\mathcal P}/\dot{q}_H$, respectively, and are shown in Figs.~\ref{2D_300K}(a) and \ref{2D_300K}(b) using the effective mass of GaAs, $m^* = 0.07 m_e$. Because of the similarity in transmission functions, the TI results are qualitatively similar to those obtained in the 1D case, except that the maximum power for the TI system occurs for $\mu > E_1$ \cite{KimR09} (see also Fig. \ref{2D_300K}(c)).  

\section{ Comparison and Discussion}

We now turn to a comparison of the three low-dimensional systems in terms of their performance under maximum power conditions. Care must be taken in a quantitative comparison of power numbers, because the QD and 1D systems produce a certain amount of power per mode (or per device), whereas the TI system has a per-area power density. Here we choose to assume a specific cross section $A_0$ for a single 1D or QD device, and express the total power output of a TI device with area $A_0$ as 
\begin{equation}
	P_{TI} = {\mathcal P_{TI}} A_0.
\end{equation}

In the following we choose $A_0$ = (10 nm)$^2$, that is, we consider, for example, an array of nanowires with one single-mode nanowire every 10 nm (a very high density \cite{Persson09}). Note that no such scaling is needed for efficiency comparisons, where cross sections drop out. 
 
\subsection{Maximum power and efficiency at maximum power}

\begin{figure}[htbp]
	\centering
	\includegraphics[scale=0.75]{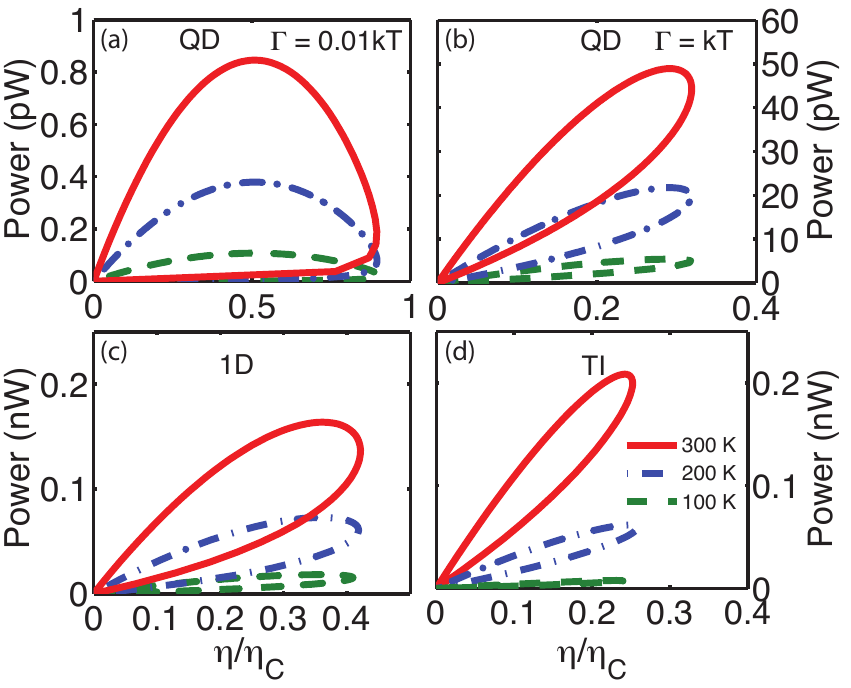}
	\caption{\label{emp loop}(Color online) (a)-(d) Loops along constant $\mu$ chosen at $P_{max}$ of each system (i.e. along the green line of Figs.~\ref{0D_300K}(b), \ref{1D_300K}(b), and \ref{2D_300K}(b)) show that efficiency at maximum power is independent of temperature. Note that the power values of the TI system depend on $A_0$ (see main text), whereas the efficiency values are independent of this choice. The QD's values depend on $\Gamma$.}
\end{figure}

In Fig.~\ref{emp loop} we show loop graphs in $(\eta, P)$ space for the value of $\mu$ that yields maximum power in each system. The horizontal position of the peak of each loop then corresponds to $\eta_{maxP}/\eta_{C}$, the normalized efficiency at maximum power. We find that, under our assumptions,  $\eta_{maxP}$ is independent of temperature in each system (for constant $\Gamma/kT$ in the case of QD). Analytically, this is because the parameters ($V, \mu$, and E) in the argument of Fermi-Dirac distribution are each scaled by thermal energy, and the integrand in the Landauer equation (Eq.~\ref{eq:current}) is constant when  $\Gamma$ is also scaled with thermal energy. Note that $\eta_{maxP}$ will certainly depend on temperature when, for example,  temperature-dependent elastic or inelastic scattering rates are taken into account. 

Quantitatively, we find that $\eta_{maxP}/\eta_{C}$ approaches the Curzon-Ahlborn value  51$\%$ for a QD system with $\Gamma << kT $ (Fig.~\ref{emp loop}(a)), as expected \cite{Esposito09} (see also Fig.~\ref{qd eff pmax vs fwhm}). For $\Gamma = 2.25 kT$ (Fig.~\ref{emp loop}(b)), where a QD system produces maximum power (see Fig.~\ref{qd eff pmax vs fwhm}), we find a drastically reduced $\eta_{maxP}/\eta_{C}$ =  17\%. The corresponding values for the 1D and TI systems are 36\% and 24\%, respectively (Figs.~\ref{emp loop}(c) and \ref{emp loop}(d)). 

The first key finding of our paper is thus that the efficiency at maximum power of a 1D system exceeds that of both the QD (at $\Gamma = 2.25kT$ where maximum power is reached, see Fig.~\ref{emp loop}(b)) and TI systems. Intuitively one may have expected this result: the main disadvantage of a QD system with finite $\Gamma$ is the low-energy tail of the transmission function, which leads to parasitic back-flow of electrons. 1D systems with their sharp onset of the transmission function at $E_1$ do not have this problem. TI systems, on the other hand, also have a sharp onset but are not actually energy filters, but filters for cross-plane momentum. Electrons with sufficient momentum to cross the barrier carry, on average, additional kinetic energy in their lateral degrees of freedom, ÒwastingÓ heat in the power-generation process \cite{Vashaee04,Vashaee04a,ODwyer05,Humphrey05}.

It should be noted that strategies exist that can improve the performance of a TI barrier by focusing electrons into the cross-plane direction \cite{Vashaee04,Vashaee04a}, using processes that lead to non-conservation of electron momentum \cite{Zide06}. In the interest of a transparent comparison of existing low-dimensional systems, such additional effects are not considered here.

\begin{figure}[htbp] 
   \centering
   \includegraphics[scale=0.75]{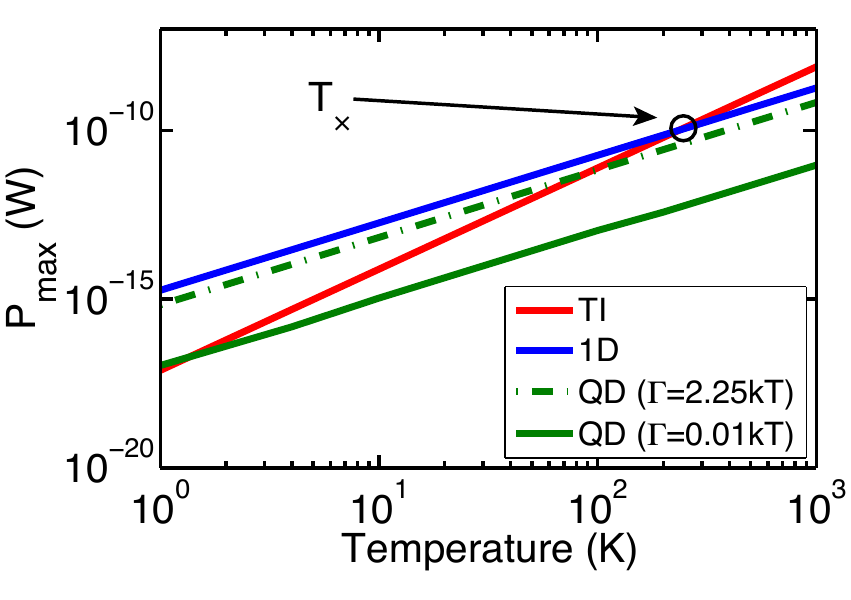} 
   \caption{\label{pmax_temp}(Color online) Maximum power as a function of temperature with $m^*=$ 0.7$m_e$ and $\Delta T/T_C =$ 0.1. $T_\times$ is the temperature where 1D and TI systems yield the same power. $T_\times$ depends on $A_0$ (see main text) and on  the electron effective mass (Eq.~\ref{TsuEsaki}. see also Fig.~\ref{t crossing}).}
\end{figure}

\begin{figure}[htbp] 
   \centering
   \includegraphics[scale=0.75]{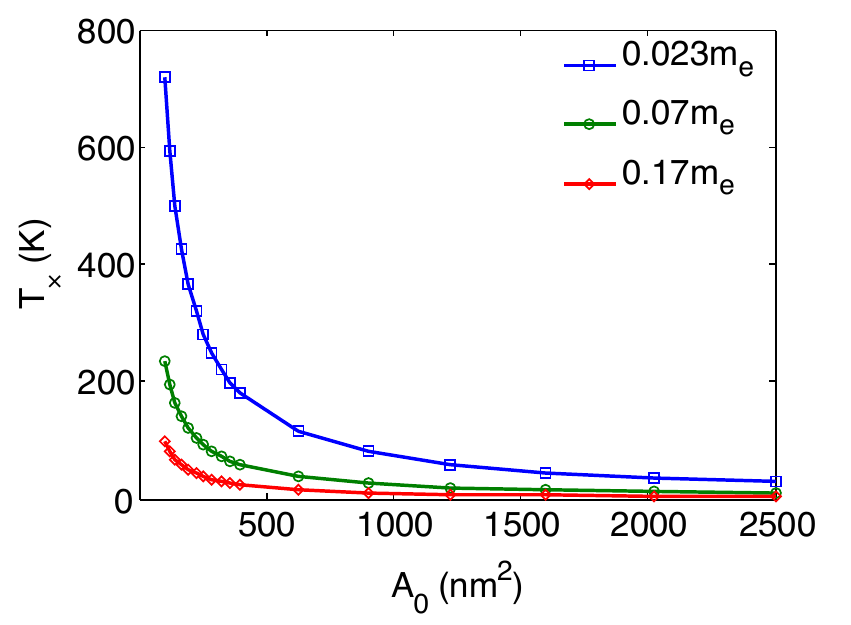} 
   \caption{ \label{t crossing}(Color online) $T_\times$ as defined in Fig. \ref{pmax_temp} as a function of effective area $A_0$ of a 1D system for different effective mass: InAs (0.023$m_e$), GaAs (0.07$m_e$), and PbTe (0.17$m_e$).}
\end{figure}

Turning now to a comparison of the absolute power values, we show in Fig.~\ref{pmax_temp} that maximum power scales with $T^2$ in the QD and 1D cases, and with $T^3$ for TI system. Note that this result is specific for a TI barrier that filters forward momentum only, and would likely be different for an ``ideal'' barrier that acts as a true energy barrier \cite{Vashaee04,Vashaee04a,ODwyer05,Humphrey05}. Overall, QD systems have the lowest power output per mode or per area. Comparing 1D and TI systems, there exists a cross-over temperature ($T_\times$), above which the TI system has the higher total power output. The value of  $T_\times$ increases with m* (because the TI power is proportional to m*) and decreases with $A_0$, because a smaller $A_0$ corresponds to a higher power density in 1D systems (Fig. \ref{t crossing}.)

\subsection{Relation to the figure of merit}

In this section we wish to place our results into the context of the traditional figure of merit for thermoelectric systems, 
\begin{equation}\label{eq:zt}
ZT=\frac{S^2\sigma}{\kappa_e+\kappa_l}T=\frac{S^2\sigma T}{\kappa_e} \left( \frac{\kappa_e}{\kappa_l+\kappa_e}\right) = (ZT)_{el} \left(\frac{\kappa_e}{\kappa_l+\kappa_e}\right)
\end{equation}
where $T $ needs to be taken as the average temperature $(T_C + T_H)/2$.  In the following we focus on $(ZT)_{el}$, the electronic part of $ZT$ as defined in Eq.~(\ref{eq:zt}), that is, we continue to assume $\kappa_l=0$. 

Unlike $P_{max}$ and $\eta_{maxP}$, $(ZT)_{el}$ is not evaluated at a specific working point with finite power output. Specifically, $S$ and $\kappa$ are defined at the open circuit condition $(I = 0)$, and $\sigma$ is evaluated at $\Delta T = 0$ in the linear response limit ($V \rightarrow 0$). We therefore  evaluate $\kappa_e$ along the open circuit curve, e.g. the red line in Fig.~\ref{0D_300K}(b). It is the ratio of $\sigma$ and $\kappa_e$ that enters $ZT$, and it can be written as
\begin{equation}
\frac{\sigma}{\kappa_e}=\frac{G}{K}=\frac{(dI/dV)_{\Delta T=0}}{(\dot{Q_H}/\Delta T)_{I=0}}.
\end{equation}
where $G$ is the electrical conductance and $K$ is the electronic thermal conductance.

\begin{figure}
	\centering
	\includegraphics[scale=0.75]{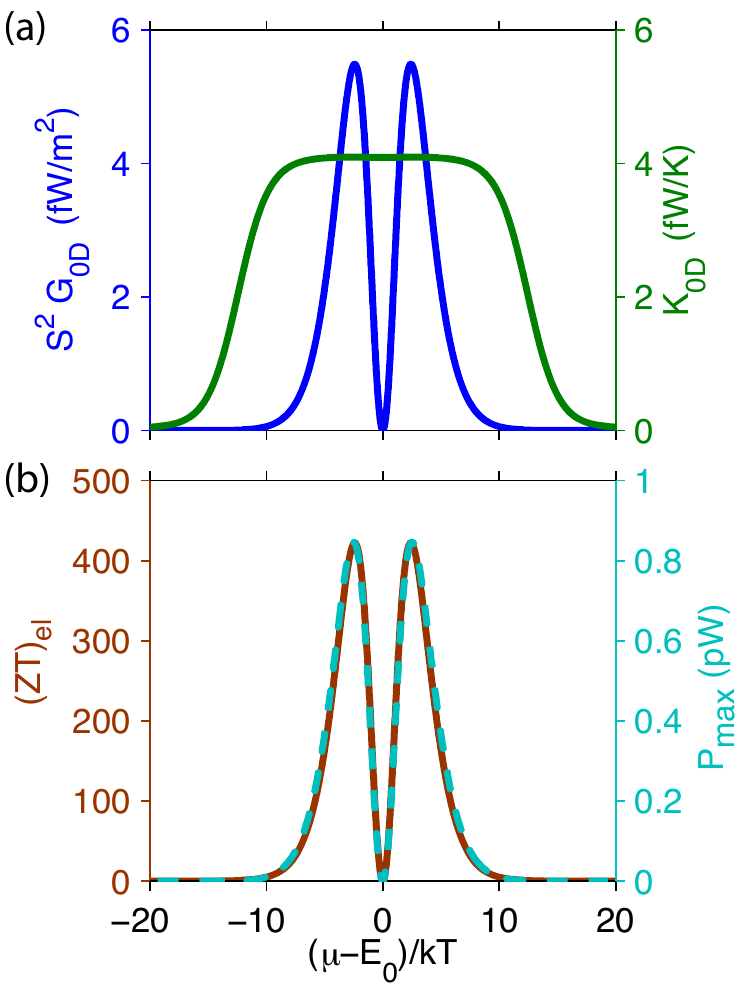}
	\caption{\label{ZT_qdot_0.01kT}(Color online) (a) Power factor $S^2 G$(blue) and thermal conductance (green), (b) $(ZT)_{el}$ (brown) and $P_{max}$ (cyan) as a function of $\mu-E_0$ for a quantum dot with $\Gamma = 0.01kT$ and $T_C=300$ K and $T_H = 330$ K. }
\end{figure}

Figure \ref{ZT_qdot_0.01kT} shows the power factor ($S^2 G$), $K$, $(ZT)_{el}$, and $P_{max}$ for a QD with a small $\Gamma=0.01kT$ at $T_C = 300$ K and $T_H = 330$ K . The power factor has a characteristic double-peak structure which results from the lineshape of a QD's Seebeck coefficient (Figs.~\ref{0D_300K}(a) and \ref{0D_300K}(b)). For small $\Gamma$, $(ZT)_{el}$ can reach enormously high values \cite{Hoffmann10}. 

\begin{figure}[htbp]
	\centering
	\includegraphics[scale=0.75]{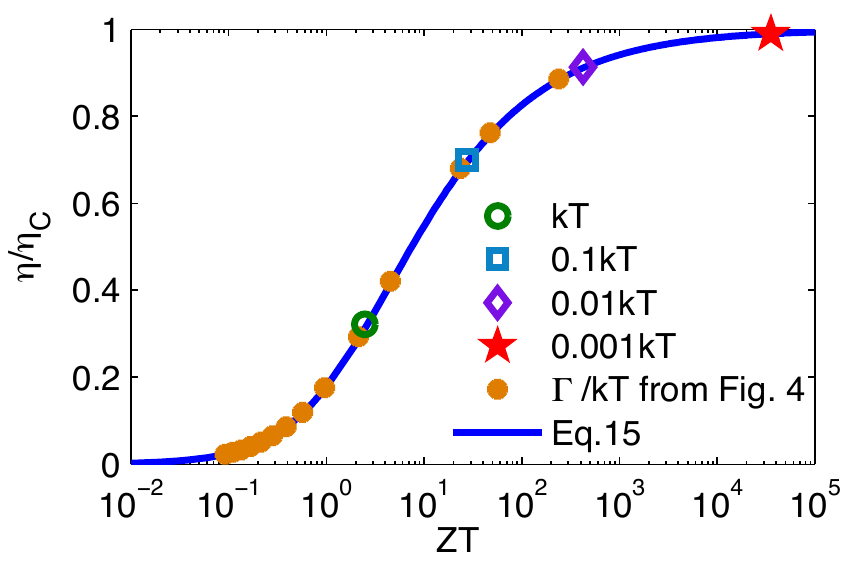} 
	\caption{\label{zt_eff}(Color online) Plot of $\eta/\eta_C$ vs. $(ZT)_{el}$ for various values of $\Gamma$ as indicated in the legend. The blue solid line is Eq.~(\ref{eff_max}).}
\end{figure}

Based on $ZT$, one can predict a thermoelectric system's maximum efficiency from \cite{Nolas01} by
\begin{equation} \label{eff_max}
\eta=\frac{M-1}{M+T_C/T_H} \eta_{C}
\end{equation}
where $M=\sqrt{1+ZT}$. In Fig.~\ref{zt_eff} we show a comparison of this equation (full line) with the exact calculations of $\eta_{max}$ (extracted from Fig.~\ref{0D_300K}(b)), graphed as a function of the maximum $(ZT)_{el}$ (Fig.~\ref{ZT_qdot_0.01kT}) for many different values of $\Gamma$, and confirm the figure of merit $ZT$ is an excellent predictor of the maximum efficiency, also in QDs. By comparison to Fig.~\ref{qd eff pmax vs fwhm},  we also find that $\eta_{max}$ (predicted by $ZT$) and $\eta_{maxP}$ agree quite well except for very small $\Gamma$, where the highest efficiency is achieved.  However, $(ZT)_{el}$ alone is clearly not a good predictor of maximum power: $P_{max}$ peaks at a value for $\Gamma$ where $(ZT)_{el}$ is well below itÕs maximum value. 

\begin{figure}
	\centering
	\includegraphics[scale=0.75]{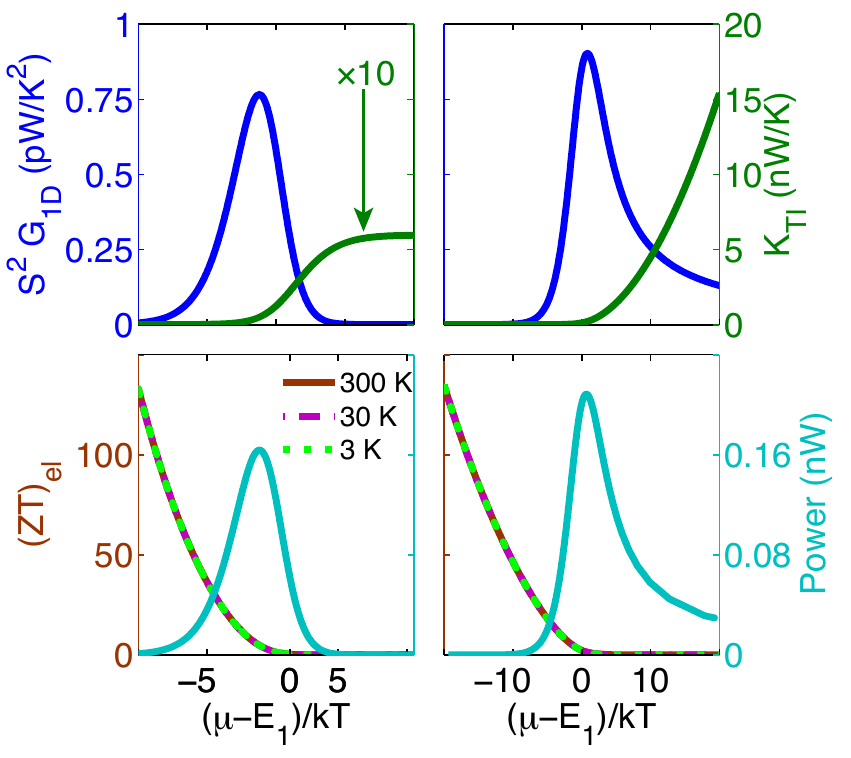}
	\caption{\label{zt 1d vs 2d}(Color online) (a) Power factor and thermal conductance, (b) $(ZT)_{el}$ and $P_{max}$ of 1D, as a function of $(\mu-E_1)/kT$ with $T_C=300$ K and $T_H=330$ K. (c) and (d) power factor, thermal conductance, $(ZT)_{el}$, and $P_{max}$ of the same condition for TI system. Note that $G$ and $K$ are obtained by assuming the effective area to be 100~nm$^2$. $(ZT)_{el}$ of 1D is  independent of temperature.}
\end{figure}

Fig.~\ref{zt 1d vs 2d} show $S^2G, K$, $(ZT)_{el}$ and $P_{max}$, for the 1D and TI systems. Note that $G$ and $K$ for the TI system have been obtained by multiplying $\sigma$ and $\kappa_e$ with $A_0$ = 100 nm$^2$. In each case, the behavior of $(ZT)_{el}$ as a function of $(\mu-E_1)/kT$ results from that of the Seebeck coefficients,  gradually increasing for $\mu$ below the band edge, and $(ZT)_{el}$ as a function of $(\mu-E_1)/kT$ is independent of temperature. Note that our value for $S^2G$ of 1D system is identical with the result in Ref.~\onlinecite{KimR09}. Like in the case of a QD, the divergence of $(ZT)_{el}$ for small $\mu$ is misleading as the power peaks for a finite value of $\mu$. 

$(ZT)_{el}$ is much higher than experimental results $ZT \approx 0.1 - 1$ for thermoelectric systems in general. However, if one includes the lattice (phonon) contribution to calculate the full $ZT$, the modeling results would reduce substantially, to those practically observed, since $\kappa_l$ is usually much higher than $\kappa_e$ in semiconductor materials. In the interest of generality, we here do not include the material-dependent $\kappa_l$ into our comparison.

\section{Conclusion}

We compared thermoelectric efficiency in the maximum power regime for a quantum dot,  a one-dimensional ballistic wire, and a thermionic power generator, each tuned to produce maximum thermoelectric power. Specifically ,we considered finite $\Gamma$ for the QD system, and found that maximum power is produced for $\Gamma = 2.25 kT$. Out of these three systems, the 1D system offers the highest achievable efficiency at maximum power, whereas above a cross-over temperature $T_\times$ a thermionic energy barrier produces the higher power per area. In our analysis we neglected the influence of parasitic heat flow carried by phonons which, in all cases, will reduce the efficiency by a factor $\dot{Q_H}/(\dot{Q}_{ph}+\dot{Q}_H)$. In a comparison between a 1D system realized using nanowires and TI systems, the influence of phonons may be to the advantage of nanowires, as surface scattering in nanowires strongly suppresses lattice heat conductivity in nanowires to a value significantly below the bulk value \cite{Mingo04,Mingo04a,Boukai08,Hochbaum08,Zhou07,Zhou10}. However, a full comparison for a specific application certainly needs to be material-specific, and should take into account the influence of a finite mean free path \cite{KimR09}, possible momentum non-conservation for electron flow across a TI barrier, which can be used to enhance the TI system's performance \cite{Vashaee04,Zide06}, and the thermoelectric properties of the material in which a TI barrier is embedded \cite {Humphrey06}. In order to place our results into context we also calculated the traditional figures of merit for each system, finding that the electronic $(ZT)_{el}$ is indeed a good predictor of the maximum electronic efficiency, but not generally for $\eta_{maxP}$. More importantly, $(ZT)_{el}$, which theoretically can be made arbitrarily large for each of the three systems studied, does not predict the working point where maximum power is achieved.

\section{Acknowledgement}

We acknowledge useful discussions with Tammy E. Humphrey, as well as financial support by a Royal Thai Government Scholarship to N.N., by Energimyndigheten, nmC@LU, ONR (N00014-05-1-0903), ARO (W911NF0720083) , by the Swedish Foundation for Strategic Research (SSF), the Swedish Research Council (VR), and the Knut and Alice Wallenberg Foundation.

\newpage
\bibliographystyle{apsrev}
\bibliography{pmax_eff}
\end{document}